# Pressure induced switching in ferroelectrics: on the junction between physics and electrochemistry

Ye Cao,[1] Anna Morozovska,[2] and Sergei V. Kalinin[1,1]

[1] The Center for Nanophase Materials Sciences, Oak Ridge National Laboratory, Oak Ridge, TN 37831

[2] Institute of Physics, National Academy of Sciences of Ukraine, 46, Prospekt Nauky, 03028 Kyiv, Ukraine

Abstract:
Pressure-induced polarization switching in ferroelectric thin films has emerged as a powerful method for domain patterning, allowing to create predefined domain patterns on free surfaces and under thin conductive top electrodes. However, the mechanisms for pressure induced polarization switching in ferroelectrics remain highly controversial, with flexoelectricity, polarization rotation and suppression, and bulk and surface electrochemical processes all being potentially relevant. Here we classify possible pressure induced switching mechanisms, perform elementary estimates, and study in depth using phase-field modelling. We show that magnitudes of these effects are remarkably close, and give rise to complex switching diagrams as a function of pressure and film thickness with non-trivial topology or switchable and non-switchable regions.

---

[1] Corresponding author, sergei2@ornl.gov

Nanoscale ferroelectrics have attracted broad attention as potential materials for domain wall electronics,[1-3] tunneling barriers,[4-6] and data storage.[7] Many of these applications are based on thin film, nanowire, or nanoparticulate materials at the device stage or intermediate synthesis steps, giving rise to significant interest in the physical functionalities and microstructure evolution in these materials. This in turn necessitates spatially resolved studies of these materials systems, readily enabled via Piezoresponse force microscopy (PFM)[8,9] and associated spectroscopies.[10-13] Multiple observations of domain nucleation and dynamics,[14-16] domain wall pinning and geometry,[17-20] and local switching behaviors[21-23] have been reported, providing new insight into physics of these materials and stimulating new directions of scientific enquiry.

The associated theory of signal formation mechanism in PFM originating from the bias induced piezoelectric deformation of the solid has been developed[8] and further extended to describe spectroscopic signals reflecting the formation and evolution of the domain or displacement of domain wall. This simple interpretation was supported by the exact[24,25] and decoupled theories[26,27] that related the measured signal, piezoelectric, dielectric, and elastic properties of materials, and domain/wall geometries,[28-30] often with the analyses available in the form of simple linear relationships. Jointly, these theoretical developments provide well developed numerical foundation of PFM, and establish the veracity of physical interpretations. This has further stimulated studies of ferroelectric switching behaviors using rigid polarization ($P = const$) models [13,31-35] and Landau-Ginzburg-Devonshire (LGD) theory based models in which polarization is defined from the minimum of corresponding free energy functional. [20,23,36-41] However, in all these analyses, the driving force for the ferroelectric switching is the modification of electrostatic energy by applied electric field, i.e., $f_{elec} \sim -\mathbf{PE}$ where $\mathbf{P}$ is the polarization vector and $\mathbf{E}$ is the electric field, and mechanical response stems purely from piezoelectricity.

While electric field induced ferroelectric switching has been well understood, there have been recent reports on mechanical pressure induced switching in ferroelectric thin film via a scanning probe. The latter becomes important as it potentially allows to create domain structures in the systems with conductive top electrode.[42] To date, pressure switching are preponderantly attributed to the flexoelectric effect, i.e. polarization and associated electric fields induced due to the large strain gradient near the tip on top of the film.[42-45] However, a broader context for tip bias and pressure -induced phenomena is given by multiple recent studies that emphasize the role of surface and electrochemical phenomena in ferroelectrics and other oxides, ranging from vacancy dynamics at surfaces and interfaces[46] to oxygen exchange to more complex electrochemical phenomena.[47,48] The possibility of tip induced electrochemical phenomena is vastly amplified under the SPM tip, with high localization of mechanical strains and electric fields. In fact, several studies exploring irreversible tip-induced electrochemistry of ferroelectrics have been reported.[47,49,50] Furthermore, chemical changes and ionic dynamics under tip underpin the electrochemical strain microscopy, demonstrated to provide readily discernible contrast to a large gamut of non-polar materials.[51,52]

These considerations necessitate the detailed analysis of the comparative role of ferroelectric, surface and bulk electrochemical phenomena in the mechanisms of PFM switching in ferroelectric films. Here, we focus on the switching under the mechanical[43] stimuli, to follow a set of work where mechanism of bias-induced strain formation in the ferroelectrics with ionic and flexoelectric couplings have been explored.[53-57] Here, we utilize both the simple analytical estimates and the phase-field modelling with chemical boundary conditions to quantify these contributions, and analyze their observability. Surprisingly, the magnitudes of these effects (given the uncertainty in experimentally known constants) are remarkably similar; suggesting the importance of multiple coupled mechanisms of pressure induced switching.

## I. Elementary phenomenology of tip-induced polarization switching

The ferroelectric materials are characterized by strong coupling between the polarization, mechanical, and chemical phenomena. Here, we classify basic mechanisms and derive simple numerical estimates for possible pressure induced switching. Here, the analysis is based on the LGD thermodynamic potential including bulk and surface contributions from ferroelectric, ferroelastic, ionic and electronic subsystems, as given in Refs [53,58,59] as well as in Suppl. Mat. along with the chosen boundary conditions and numerical values of parameters collected from the Refs [60-62]. In this, Euler-Lagrange equations for ferroelectric polarization components $P_i$ obtained from the LGD-potential variation are coupled with electrostatic equations for electric field, material equations relating the field and displacement, generalized Hooke's relations and mechanical equilibrium equations for elastic stresses $\sigma_{ij}$ and strains $u_{ij}$. Elastic boundary conditions are $\sigma_{ij} n_j \big|_{S_f} = -p_i^{ext}$ at the free surfaces of the system, where $n_j$ is the component of the outer normal to the surface, $p_i^{ext}$ is the external pressure (e.g. imposed by the tip). The boundary conditions for $P_i$, which in general case should be of the third kind, [63] become inhomogeneous via joint action of external pressure and the flexoelectric coupling,[59] namely $\left( \frac{F_{kl33}}{g_{33}^{eff}} \sigma_{kl} \mp \frac{P_3}{\lambda} - \frac{\partial P_3}{\partial x_3} \right)\bigg|_{x_3=0,h} = 0$. $F_{kl33}$ is the component of the flexoelectric effect tensor. The extrapolation length $\lambda$ is equal to the ratio of surface energy coefficient $A_{33}^S$ and the nonzero effective polarization gradient coefficient $g_{33}^{eff}$, $\lambda = g_{33}^{eff}/A_{33}^S$. Realistic range for $\lambda$ is 0.5 – 50 nm,[59,64] but the phenomenological parameters is usually unknown (as defined by the surface energy and short-range interactions [65]). Generally, it is beneficial to consider the two limiting cases of very small and big $\lambda$, with the latter corresponding to the minimal contribution from the flexoeffect near the surface. However, such an analysis necessarily lacks transparency, and calls for numerical estimates.

Here, we enumerate main mechanisms that can be derived from LGD functional or derived based on elementary physical considerations.

**I.a. Ferroelastic mechanism:** The application of the mechanical pressure to the SPM tip creates the driving force for the ferroelastic switching between mechanically-non-equivalent structural variants. In comparison, it does not induce switching between antiparallel domains, i.e. classical ferroelectric switching. Notably, for unclamped crystal the switching will be thermodynamically favorable for arbitrarily small pressure. For clamped crystal the normal/lateral polarization is enhanced/inhibited, and a larger pressure is required for the in-plane switching. Based on LGD theory normal polarization suppression could even occur in clamped crystal before the in-plane switching. [Figure 1(a)] Hence establishing the critical pressure for switching from out-of-plane to in-plane domain configuration for material with rigid polarization (i.e. **P** = const) necessitates analysis of nucleation and clamping effects to derive switching fields, etc.

**I.b. Polarization suppression.** The application of the pressure can suppress polarization in a non-rigid multiaxial ferroelectric with coordinate-dependent polarization vector $\mathbf{P}(\mathbf{r},t)$. To derive the estimates of corresponding critical pressure, we note that the internal stresses $\sigma_{ij}$ couple to the polarization direction via piezoelectric and electrostriction couplings.[61,66] Within LGD theory, the piezoelectricity is described as linearized electrostriction, and corresponding effective piezoelectric coefficient is $d_{ijk} = 2\varepsilon_0 (\varepsilon_{km}^f - \delta_{km}) Q_{ijml} P_l^S$, where $Q_{ijkl}$ is electrostriction tensor, $P_k^S$ is a spontaneous polarization component, $\varepsilon_0$ is the dielectric permittivity of vacuum, $\delta_{km}$ is a Kroneker symbol and $\varepsilon_{ij}^f$ is the relative dielectric permittivity of ferroelectric that includes a soft-mode related electric field-dependent contribution $\varepsilon_{ij}^{sm}$. The electrostrictive coupling renormalizes the coefficients in LGD-thermodynamic potential (see Suppl. Mat.). Namely, the coefficient $a_{33}$, that determines the FE transition temperature of the out-of-plane polarization component, becomes $a_{33}^{eff} = \alpha_T (T - T_c) + 2 Q_{33}^{eff} p_{ext}$, where $Q_{33}^{eff} = Q_{33} - 2 s_{12} Q_{12} / (s_{11} + s_{12})$ and so the last term increases or decreases $a_{33}^{eff}$ depending on the $p_{ext}$ sign. The coefficient $a_{11}$, that determines the ferroelectric transition temperature of the in-plane polarization component, becomes $a_{11}^{eff} = \alpha_T (T - T_c) + Q_{11}^{eff} p_{ext}$, where $Q_{11}^{eff} = (Q_{12} s_{11} - Q_{11} s_{12}) / (s_{11} + s_{12})$ and $s_{ijkl}$ is elastic stiffness coefficient.

Here, we perform the estimates for PbZr$_{0.2}$Ti$_{0.8}$O$_3$, **PZT(20/80)**, for which $Q_{33}^{eff} \cong 0.4307$ C$^{-2}$ m$^4$ and $Q_{11}^{eff} \cong +0.0036$ C$^{-2}$ m$^4$. Since $Q_{33}^{eff} >> Q_{11}^{eff} > 0$, the negative pressures (**compression**) more strongly favors out-of-plane polarization component with the spontaneous value $P_3[p_{ext}] \cong \sqrt{-a_{33}^{eff}[p_{ext}]/a_{111}}$. At that a renormalized temperature of PE phase instability is $T_3^c[p_{ext}] = T_c - 2 Q_{33}^{eff} p_{ext} / \alpha_T$. Compression slightly increases the in-plane component so that $P_1[p_{ext}] \cong \sqrt{-a_{11}^{eff}[p_{ext}]/a_{111}}$ and corresponding transition temperature $T_1^c[p_{ext}] = T_c - Q_{11}^{eff} p_{ext} / \alpha_T$, where $\alpha_T = 3.12 \times 10^5$ C$^{-2}$ Jm/K and $T_c = 768$ K. At the same time, positive pressures (**tension**) strongly suppress the out-of-plane component appearance, and slightly suppress the in-plane one. The estimates are in qualitative agreement with the results. [61,67]

The 1D estimate of the maximal normal pressure corresponding to the tip-surface mechanical contact radius $R = 50$ nm and indentation force $N = 1$ μN, gives the pressure under the tip as $|p_i^{ext}| \cong N/(\pi R^2) \approx 10^{-6}/(25\pi \times 10^{-16}) = 1.27 \times 10^8$ Pa. The pressure favours the out-of-plane polarization, since $T_3^c[p_{ext}] \approx 1110$ K is about increased by 342 K in comparison with $T_c = 768$ K. Hence, within the 1D estimate $P_3[p_{ext}]$ becomes 1.45 times more at room temperature. At the same time in-plane component changes relatively insufficiently and $T_1^c[p_{ext}]$ shift at about 1K, because $Q_{11}^{eff}$ is about 100 times smaller than $Q_{11}^{eff}$. To resume, the effect of pressure (compression $p_{ext} < 0$ or tension $p_{ext} > 0$) on $T_3^c[p_{ext}]$ depends on the signs and relative magnitude of $Q_{11}^{eff}$ and $Q_{33}^{eff}$. The pressure increases the out-of-plane $T_3^c[p_{ext}]$ if $-Q_{33}^{eff} p_{ext} > 0$ and decreases it in the materials with $-Q_{33}^{eff} p_{ext} < 0$.

The situation becomes more complex for the SPM geometry. In this case, the tip-induced pressure is concentrated directly below the tip and is sensitive to the exact tip geometry. For rigid piezoelectric materials, the corresponding field distributions have been calculated.[24,25,68-71] Similarly to classical Hertzian indentation, the pressure (and electric field) is maximal at the periphery of the contact area and decreases away from it, and both in-plane and out of plane components are present. Under the conditions inducing switching, the concentration of field below tip serve as potential nucleation centers, with the type of switching process being determined by the interplay between materials and tip field symmetry. For example, for the c domain in tetragonal material, some of the possible domain configurations are shown in Figure 1(b). Note that some of these will be associated with the charged domain walls under the tip; others will be charge neutral under the tip and generate charges at the boundary forming dipolar field. The direction of symmetry breaking will be affected by local defects, tip motion,[39] etc. However, it is important to note that we can expect significant changes in induced ferroelastic domains structures under pressure and when pressure is reduced, and strong dependence on defects and trajectory of tip motion. While ferroelectric domain switching is possible in this case, it will be the result of interplay of multiple secondary mechanisms and cannot be predicted without phase field modelling with defined surface conditions.

**I.c. Flexoelectric mechanisms.** The next broadly considered mechanism is flexoelectric switching, suggested by Gruverman et. al.[43] By definition, stress gradient in the tip-surface junction generates built-in field in the material that can result in switching. As it was shown in ref [59] the main contribution of the flexoelectric effect is the appearance of the out-of-plane surface polarization $P_3^{FL}$ and built-in normal flexoelectric field $E_3^{FL}$, which are proportional to the product $F_{33} p_i^{ext}$. For nonzero extrapolation length $\lambda$ it can be estimated as

$P_3^{FL} \sim \dfrac{h\lambda F_{33}^{eff} \sigma_{33}}{g_{33}^{eff}(h+\lambda)} \cong \dfrac{F_{33}^{eff} p_i^{ext} h\lambda}{g_{33}^{eff}(h+\lambda)}$, where effective flexocoefficient is $F_{33}^{eff} = F_{33} - 2s_{13}F_{13}/(s_{11}+s_{12})$, and $h$ is a slab thickness that can be substituted by the characteristic depth of the indentation (for the latter case $h \sim R$). The estimates of the maximal normal pressure corresponding to the tip-surface contact radius $R = 50$ nm and identation force $N = 1$ μN, gives the high enough pressure $p_i^{ext} \cong 1.27 \times 10^8$ Pa. The surface polarization $P_3^{FL}$ estimated for typical flexocoefficient range $F_{33}^{eff} = (0.5 - 5) \times 10^{-11}$C$^{-1}$m$^3$, gradient coefficient $g_{33}^{eff} = (0.5 - 2) \times 10^{-10}$C$^{-2}$m$^4$N,[72] $h=10$ nm and $\lambda \gg h$ gives $P_3^{FL} = (0.015 - 1.2)$ C/m$^2$; so that it can be small enough but also can be comparable with the spontaneous polarization value, or even several times higher. The values in the middle of this range are sufficient to stimulate polarization switching of in the bulk. Corresponding flexoelectric field is inversely proportional to the thickness $h$, $E_i^{FL} \sim F_{ijkl} \partial \sigma_{jk}/\partial x_l \sim F_{33} p_i^{ext}/h$. Thus for the case $h \sim R$ it gives $E_3^{FL} \sim 0.5$ V/nm or even higher. Note that the latter interval contains coercive field, and so can indeed reverse the polarization. However, being rather important qualitatively, the above 1D estimations of $P_3^{FL}$ and $E_3^{FL}$ can be of small significance quantitatively to the great discrepancy (several orders of magnitude) in the numbers steaming from the scattering for the flexoelectric tensor $F_{ijkl}$ and gradient tensor $g_{ij}^{eff}$ values (several orders of magnitude) in the literature even for well-known ferroelectrics.[73-75] A possible escape from the situation is to decouple the problem, i.e. to consider all contributions separately in a rigorous phase-field modelling suggesting coupled 3D problem. To realize the idea one can consider zero or negligibly small convolution $F_{ijkl}^{eff} \sigma_{kl}$ at the surface as well as $\left.\dfrac{\partial P_3}{\partial x_3}\right|_{x_3=0,h} = 0$ that will be done in the next section.

However, this simple estimate has to be complemented by several important considerations. First of all, the flexoelectric field is active only in the pressure on state, and disappears instantly when the pressure is removed. Correspondingly, it can facilitate polarization switching via complementary fields (e.g. pressure assisted bias-induced switching detectable as lateral shift of hysteresis loops) and result in metastable domain configurations, but is not equivalent to uniform electric field effect. Secondly, the flexoelectric induced strain gradient and resultant electric field are subject to the boundary conditions at the tip and in the bulk [Figure 2(b)], and can have complex dipolar structure for the grounded tip [Figure 2(c)]. Thus the polarization switching could be strongly size-dependent [Fig. 2(d)]. Finally, similarly to ferroelastic effects, flexoelectric field has both in-plane and out of plane components, potentially leading to complex metastable domains structures sensitive to load history and tip motion.

**I.d. Surface electrochemistry.** Pressure will strongly affect the electrostatics of screening charges on ferroelectric surfaces, hence affecting the stability of associated polarization state. For example, the positive polarization state screened by OH- and negative polarization charge screened by H+ will respond differently to external pressure, with the chemical shift defined by the difference of molar volume between the two. The normal component of the effective electric field induced by the surface ions with a Langmuir-type charge density $\sigma$ is given by expression,[76] $E_3^{SEC}(U,\sigma) = \frac{\varepsilon_0 \varepsilon_d U + \Lambda \sigma}{\varepsilon_0 (\varepsilon_d h + \Lambda \varepsilon_{33}^b)}$, where $\Lambda$ is the effective separation between the ferroelectric polarization and the ion layer, $\varepsilon_d$ is the relative dielectric permittivity of the separated region (i.e. it is a background permittivity in the considered case, $\varepsilon_d = \varepsilon_{33}^b \sim 5$). Since $\lambda$ is about or smaller than the lattice constant (0.4 nm), and $h$ has an order of the contact radius R, the inequality $h \gg \Lambda$ is valid, and so $E_3^{SEC}(U,\sigma) \approx \frac{\varepsilon_0 \varepsilon_{33}^b U + \Lambda \sigma}{\varepsilon_0 \varepsilon_{33}^b h}$. Estimates at zero applied bias (U=0) gives $E_3^{SEC}(\sigma) \approx \Lambda \sigma / (\varepsilon_0 \varepsilon_{33}^b h) \sim \Lambda P_S / (\varepsilon_0 \varepsilon_{33}^b h) \sim \frac{0.5 nm}{5 nm} \frac{0.5 C/m^2}{5 \cdot 8.85 \times 10^{-12} F/m} \sim 10^9$ V/m $\sim$ 1 V/nm. As one can see the surface electrochemical contribution in number can be comparable with the flexoelectric one, but it tends to zero at $\Lambda \to 0$. Note that both contributions are proportional to $1/h$.

Unlike other coupling mechanisms, the relationships between applied pressure and electrochemical potential shift can be highly non-monotonic. For example, for water the difference in molar volume for $H_3O+$, OH- and H+ can give rise to oscillatory dependence of chemical potential on pressure, similar to phenomena observed in ionic liquids.[77-79] Finally, relevant consideration here is that of triboelectric charging, known to produce very high (tens and hundreds of V) potentials (but quantitative mechanisms of which are not well-known).

**I.e. Bulk electrochemistry.** Finally, applied pressure can shift electrochemical potential of vacancies, resulting in local charging and resultant polarization switching. In this case, the relationship between applied pressure and induced potential potentials is most complex, since the coupling proceeds through multiple steps affected by relevant boundary conditions, etc. Notably, for surface and bulk electrochemical mechanisms time becomes a significant component. For example, the development of potential distributions in electrochemical systems in response to step wise change in external stimulus (e.g. pressure or bias at the electrode) is determined by a set of relaxation times, i.e., the Debye time ($\tau_D$), the bulk diffusion time ($\tau_L$) and the diffuse charge relaxation time ($\tau_C$) from the physics of diffuse-charge dynamics.[80,81] The ionic migration and accumulation/depletion in electrochemical system under applied field forms electrical double layers that can screen the electrode. The Debye time describes the characteristic time for ion to diffuse across electric double layer while the bulk diffusion time is for ion diffusion from neutral bulk towards the electrodes. They are determined by $\tau_D = \lambda_D^2/D$ and $\tau_L = L^2/D$ respectively where $\lambda_D$ and $L$ are the Debye screening length and the separation distance between electrodes, and $D$ is the ionic diffusivity. The Debye screening length is determined from $\lambda_D = \sqrt{(\varepsilon k_B T)/(2z^2 e_0^2 C)}$ where $C$ is the ionic concentration, $e_0$ is the electronic charge, $z$ is the charge number, $k_B$ is the Boltzmann constant, $T$ is the temperature and $\varepsilon$ is the permittivity of the host material. For oxygen vacancies at a concentration of $10^{18}$ cm$^{-3}$ with a diffusivity of $\sim 10^{-16}$ cm$^2$/s at room temperature[82] in 10nm thick PZT thin film, the Debye length Debye time and bulk diffusion time are estimated to be 0.422nm, 6.37s and 3580s. $\tau_D$ is deterministic for surface electrochemical behavior and $\tau_L$ for bulk diffusion mechanism. For diffuse charge relaxation dynamics, a third primary time scale is for ($\tau_C$) is introduced as the harmonic mean of Debye and bulk diffusion time $\tau_C = \sqrt{\tau_D \tau_C} = \lambda_D L/D$, which is estimated to be 151.1s.

## II. Phase field modelling

The analysis above provides simple estimates of the pressure induced phenomena in the ferroelectrics. However, these mechanisms are intrinsically coupled, and in many cases the indirect couplings (e.g. pressure -> vacancy concentration -> electric field -> polarization dynamics) can exceed direct coupling. Hence, we analyze the phenomena using coupled phase field model by considering the surface chemical effect, flexoelectric effect and bulk Vegard strain effect, as schematically illustrated in Figure 3.

### II.a. Description of phase field model

In the phase-field simulations of ferroelectric phenomena, we choose ferroelectric polarization ($P_i$, i=1~3) as the order parameter, and the total free energy of the system is written as a function of $P_i$, elastic strain ($\varepsilon_{kl}$), electric field ($E_i$) and the polarization gradient ($\nabla P_i$).

$$f = f_{land}(P_i) + f_{elas}(P_i, \varepsilon_{kl}) + f_{elec}(P_i, E_i) + f_{grad}(\nabla P_i) \tag{1}$$

in which $f_{land}$, $f_{elas}$, $f_{elec}$ and $f_{grad}$ represent the Landau bulk free energy density, the elastic energy density, the electrostatic energy density and gradient energy density respectively. $f_{land}$ is written as a sixth order polynomial expansion of $P_i$.

$$\begin{aligned} f_{land}(P_i) &= \alpha_i P_i^2 + \alpha_{ij} P_i^2 P_j^2 + \alpha_{ijk} P_i^2 P_j^2 P_k^2 \\ &= \alpha_1(T)\left(P_1^2 + P_2^2 + P_3^2\right) + \alpha_{11}\left(P_1^4 + P_2^4 + P_3^4\right) + \alpha_{12}\left(P_1^2 P_2^2 + P_2^2 P_3^2 + P_3^2 P_1^2\right) \\ &\quad + \alpha_{111}\left(P_1^6 + P_2^6 + P_3^6\right) + \alpha_{112}\left[P_1^2\left(P_2^4 + P_3^4\right) + P_2^2\left(P_1^4 + P_3^4\right) + P_3^2\left(P_1^4 + P_2^4\right)\right] + \alpha_{123} P_1^2 P_2^2 P_3^2 \end{aligned} \tag{2}$$

where $\alpha_i, \alpha_{ij}$ and $\alpha_{ijk}$ are the second, fourth and sixth order Landau coefficients. Only $\alpha_1$ is linearly dependent on temperature $\alpha_1 = \dfrac{1}{2\varepsilon_0 C}(T - T_0)$. Here $\varepsilon_0$ is the vacuum permittivity, $C$ is the Curie constant and $T_0$ is the transition temperature.

The elastic energy density is written as,

$$f_{elas} = \frac{1}{2} c_{ijkl}(\varepsilon_{ij} - \varepsilon_{ij}^0)(\varepsilon_{kl} - \varepsilon_{kl}^0) \tag{3}$$

in which $c_{ijkl}$ is the elastic stiffness coefficient tensor, $\varepsilon_{ij}$ is the total strain and $\varepsilon_{ij}^0$ is the eigenstrain. In the absence of ionic defects such as oxygen vacancies, $\varepsilon_{ij}^0$ is induced by the spontaneous polarization as,

$$\varepsilon_{ij}^0 = \varepsilon_{ij}^{0P} = Q_{ijkl} P_k P_l \qquad (4)$$

in which $Q_{ijkl}$ is the electrostrictive coefficient tensor. The electrostatic energy of a domain structure is introduced through,

$$f_{elec}(P_i, E_i) = -P_i E_i - \frac{1}{2}\varepsilon_0 \varepsilon_r E_i E_j \qquad (5)$$

where $E_i$ is the total electric field. $\varepsilon_0$ and $\varepsilon_r$ are the vacuum permittivity and relative permittivity respectively. $E_i$ is related to the electric potential ($\phi$) distribution through,

$$E_i = -\nabla_i \phi \quad (i = 1 \sim 3) \qquad (6)$$

where $\nabla$ is the gradient operator. The gradient energy density is introduced through the polarization gradient,

$$f_{grad}(\nabla P_i) = \frac{1}{2} g_{ijkl} \left( \frac{\partial P_i}{\partial x_j} \frac{\partial P_k}{\partial x_l} \right) \qquad (7)$$

in which $g_{ijkl}$ are the gradient energy coefficient tensor.

The temporal evolution of ferroelectric polarization is governed by the time-dependent Landau-Ginzburg-Devonshire (LGD) equations,

$$\frac{\partial P_i(\boldsymbol{x}, t)}{\partial t} = -L \frac{\delta F_{total}}{\delta P_i(\boldsymbol{x}, t)}, i = 1, 2, 3 \qquad (8)$$

in which $\boldsymbol{x}$ is the position, $t$ is the time, $L$ is the kinetic coefficient related to the domain movement, $\delta$ is the variational derivative operator, and $F_{total} = \int_V f dV$ is the total free energy written as the volume integral of $f$.

To model the effect of mechanical pressure on the ferroelectric thin film induced from a scanning probe, we define the stress distribution as that created by a spherical indenter on top of the film. Using Hertzian model for isotropic solid,

$$\sigma_{33}^{tip}(r) = \begin{cases} -\dfrac{3p}{2\pi a^2}\sqrt{1-\dfrac{r^2}{a^2}} & (r \leq a) \\ 0.0 & (r \geq a) \end{cases} \quad (9)$$

where $p$ is the mechanical load, $a$ is the radius of the tip-surface contact area, and $r = \sqrt{(x-x_0)^2 + (y-y_0)^2}$ is the distance from any points $(x, y)$ inside the contact area to the tip center $(x_0, y_0)$.

The local mechanical tip pressure would create inhomogeneous strain, and consequently a large strain gradient near the tip in nanoscale thin film. While homogeneous strains ($\varepsilon_{kl}$) would induce polarizations in ferroelectric oxides through the piezoelectric effect, the inhomogeneous strain or strain gradient ($\partial \varepsilon_{kl}/\partial x_j$) would additionally contribute to the ferroelectric polarization through the flexoelectric effect, i.e.,

$$P_i = d_{ijk}\varepsilon_{jk} + \mu_{ijkl}\frac{\partial \varepsilon_{kl}}{\partial x_j} \quad (i,j,k,l = 1 \sim 3) \quad (10)$$

where $d_{ijk}$ and $\mu_{ijkl}$ are the piezoelectric and flexoelectric polarization tensors respectively. The coupling between $P_i$ and $\partial \varepsilon_{kl}/\partial x_j$ give rises to an additional flexoelectric energy density ($f_{\text{flexo}}$) to be added to the total free energy density defined in Eq. (1),

$$f_{\text{flexo}}(P_i, \varepsilon_{kl}, \nabla P_i, \nabla \varepsilon_{kl}) = \frac{1}{2}f_{ijkl}\left(\frac{\partial P_k}{\partial x_l}\varepsilon_{ij} - \frac{\partial \varepsilon_{ij}}{\partial x_l}P_k\right) = \frac{1}{2}F_{ijkl}\left(\frac{\partial P_k}{\partial x_l}\sigma_{ij} - \frac{\partial \sigma_{ij}}{\partial x_l}P_k\right) \quad (11)$$

in which $f_{ijkl}$ (unit: V) and $F_{ijkl}$ (unit: Vm$^2$N$^{-1}$) are the flexocoupling coefficient tensors. The relations between $f_{ijkl}$, $F_{ijkl}$ and $\mu_{ijkl}$ are $f_{ijkl} = c_{ijmn}F_{mnkl}$, $\mu_{ijkl} = \varepsilon_0 \chi_{mn} f_{mnkl}$ where $\chi_{mn}$ is the dielectric susceptibility.

The presence of charged defects such as oxygen vacancies in ferroelectric oxide thin film create additional charge compensation and strain relaxation, both of which can play an important role in the ferroelectric switching behavior. Similarly, the polarization distribution induces local bound charges and spontaneous strains that will affect the oxygen vacancy transport under mechanical tip pressure. To consider this coupled effect, we solve the electrostatic equilibrium (Poisson) equation for the electric potential ($\phi$) distribution,

$$-\nabla^2 \phi = \frac{z_i e_0 c_i - \nabla \cdot P_i}{\varepsilon_0 \varepsilon_r} \quad (12)$$

in which $z_i$ is the charge number for species $i$, $e_0$ is the unit charge and $c_i$ is the concentration of charge species $i$. The local eigenstrain ($\varepsilon_{ij}^{0d}$) induced from the oxygen vacancy is described by the converse Vegard effect,

$$\varepsilon_{ij}^{0d} = V_{ij}^d \Delta X_V \delta_{ij} \tag{13}$$

where $\delta_{ij}$ is the Kroneker operator, $V_{ij}^d$ is the Vegard coefficient which measures the change of lattice parameter ($a$) with respect to oxygen vacancy composition ($X_V$), i.e., $V_{ij}^d = (1/a)(da/dX_V)$. $\Delta X_V = X_V - X_{V0}$ is the variation of ionized oxygen vacancy composition where constant value of $X_{V0}$ corresponds to a stress-free reference state at zero electric field. The unitless oxygen vacancy composition is related to its concentration ($[V_O^{\bullet\bullet}]$, in the unit of #/cm$^3$) through,

$$X_V = \left([V_O^{\bullet\bullet}] \bullet AW\right)/\left(\rho \bullet N_A\right) \tag{14}$$

where $N_A = 6.022 \times 10^{23}$ (#/mol) is the Avogadro constant. AW and $\rho$ represent the atomic weight and density of the matrix material. For PZT we choose AW=303.09 (g/mol) and $\rho$=7.52 (g/cm$^3$). Thus Eq. (4) should be modified as,

$$\varepsilon_{ij}^0 = \varepsilon_{ij}^{0P} + \varepsilon_{ij}^{0d} = Q_{ijkl} P_k P_l + V_{ij}^d \Delta X_d \delta_{ij} \tag{15}$$

To describe the oxygen vacancy transport driven by the mechanical pressure in the presence of local charge and strain distribution, we solve the Nernst-Planck transport equation for oxygen vacancy,

$$\frac{\partial X_V}{\partial t} = -\nabla \bullet J = D_V \nabla^2 X_V + \frac{D_V z_V e}{k_B T} \nabla \bullet [X_V \nabla \phi] + \frac{D_V}{RT} \nabla \bullet [X_V \nabla \mu_{el}] \tag{16}$$

where $t$ is the time step, $J$ is the flux of concentration, $D_V$ is the diffusivity of oxygen vacancy, $k_B$ is the Boltzmann constant, $R$ is the gas constant, and $T$ is the temperature. The elastic potential ($\mu_{el}$) is written as,

$$\mu_{el} = V_{ii}^d \sigma_{ii} = V_{11}^d \sigma_{11} + V_{22}^d \sigma_{22} + V_{33}^d \sigma_{33} \tag{17}$$

Detailed derivation of Eq. (16) and (17) can be found in Suppl. Mat..

To study surface electrochemical effect on the mechanical switching dynamics, we applied phase-field model with chemical boundary condition.[83] In classic phase-field model, the electrostatic equilibrium equation [Eq. (12)] and LGD equations [Eq. (8)] are typically solved with boundary conditions,

$$\phi|_{z=0} = 0 \text{, and } \phi|_{z=L} = V_{app} \qquad (18)$$

$$\left.\frac{\partial P_z}{\partial z}\right|_{z=0,L} = 0 \qquad (19)$$

in which $L$ is the film thickness and $V_{app}$ denotes the applied electric bias. These represent the situation when the polarization bound charges at the top/bottom surface are fully screened by the metal electrode at fixed bias ($V_{app}$). While these boundary conditions match well with the bulk ferroelectrics, they are not applicable to free ferroelectric surfaces. Here we used chemical boundary condition based on S&H model,[84] which maintains electrochemical equilibria between surface compensating charges and electrochemical potentials and environment. For simplicity, we consider that surface charges include only excessive oxygen ions (such as negatively charged adsorbed oxygen) and deficient oxygen ions (such as positively charged oxygen vacancy). The surface charges are assumed to reside in the dielectric layer of thickness $\lambda$ atop the ferroelectric thin film, where ferroelectric polarization vanishes. Thus Eq. (18) and (19) are modified as,

$$\phi|_{z=0} = 0 \text{, and } \phi|_{z=L+\lambda} = V_{ex} \qquad (20)$$

$$P_z|_{z=L+\lambda} = 0 \text{, and } \left.\frac{\partial P_z}{\partial z}\right|_{z=0} = 0 \qquad (21)$$

$V_{ex}$ is the electrochemical potential which is related to the surface ion concentration ($\theta_i$), oxygen partial pressure ($P_{O_2}$) and standard free energy of surface ion formation ($\Delta G°$) through the Langmuir adsorption isotherm,

$$\frac{\theta_i}{1-\theta_i} = P_{O_2}^{1/n_i} \exp\left(\frac{-\Delta G° - z_i e_0 V_{ex}}{k_B T}\right) \qquad (22)$$

More detailed formulation of chemical boundary condition can be found in literature.[83,84] The electrochemical potential of the screening surface ions is also subjected to the mechanical pressure at the tip-surface junction. The shift of formation energy ($\Delta G°$) by the tip pressure can be estimated to be $P\Delta V$, where $P$ is the applied pressure and $\Delta V$ is the difference in ionic volume. Therefore, at fixed $\theta_i$ and $P_{O_2}$, $V_{ex}$ varies at different tip pressures.

In the next sections, we will systematically analyze the surface chemical effect, flexoelectric effect and bulk electrochemical effect on the mechanical switching in ferroelectric thin film at different thickness under different tip pressures. Based on this we will elucidate their relative contributions and couplings in the switching dynamics. We chose Pb($Zr_{0.2}Ti_{0.8}$)$O_3$ (PZT) as an example, and the materials parameter and simulation conditions are listed in Table I in the Suppl. Mat..

**II.b. Ferroelastic phenomena**

We started with the polarization state under pure mechanical pressure from the probing tip without considering the surface charge effect, the flexoelectric effect and the bulk vacancy transport effect. One of the advantages in phase-field method is that one can easily separate contributions from different effects and understand their relative roles in the mechanical switching. To do this we chose conventional boundary conditions for polarization and electric potential [see Eq. (18) and (19)], set the flexoelectric coefficients to be 0, assumed charge neutral in the bulk and froze the oxygen vacancy migration. We applied tip load from 0.5μN to 4.0μN on PZT thin film consisting of a (001) oriented single domain, with thickness from 5nm to 20nm. Figure 4 (a - d) illustrate the final polarization states under lower (0.5μN) and higher (4.0μN) mechanical loads in thinner (5.0nm) and thicker (20nm) films. For all thickness, the magnitude of $P_z$ component near the tip was slightly suppressed from 0.8C/m$^2$ under 0.5μN, and eventually became 0.0C/m$^2$ under 1.0μN (Figure S1), indicating that the polarization switched to in-plane orientation beneath the tip. The in-plane switching regions eventually penetrate through the entire film depending on the film thickness. Notably no 180° switching occurred even when tip load increased to 4.0μN, implying that the pure ferroelastic effect is symmetric. This is further evidenced by the almost symmetric distribution of out-of-plane electric field ($E_z$) under the tip region, [Figure 4(e)], where the upward $E_z$ near the bottom layer prevented the in-plane polarization from further switching into [00$\bar{1}$] orientation.

**II.c. Surface electrochemistry**

To study the surface charge effect on the switching dynamics, we applied chemical boundary conditions [Eq. (20) ~ (22)] taking into account the interaction between surface charge concentration, electrochemical potential and oxygen partial pressure. The flexoelectric effect and bulk vacancy effect were turned off. Figure 5 (a - d) show the final polarization states under different tip loads in PZT thin films of different thickness. Clearly 180°switching were seen in ultrathin film (5.0nm) under 0.5μN. This is due to the shift of electrochemical potential on top of the film. The electrochemical potential shift scales with tip pressure, creating an additional downward electric field under the tip (Figure 5(e)). When the tip load increased to 4.0μN, the 180°switching region was mainly seen at the edges. Inside the tip region the polarization became in-plane [Figure 5(b)]. This implies that there is a competition between the electrochemical shift that favors the 180°switching, and the pure ferroelastic effect that favors in-plane switching (as studied in II.b), and both effects increase with tip pressure. This is clearly illustrated in 10nm thin film (see Figure S2) where polarization experienced three distinctive states (suppression, 180° switching and in-plane switching) with increasing tip pressures. Below 10nm surface electrochemical effect is dominant, while above 10nm ferroelastic effect takes over. Therefore 180°switching only occurred in ultrathin film below 10nm when surface effect is on.

**II.d. Flexoelectric effect**

The high localization of mechanical strains from the tip load couples with ferroelectric polarization through flexoelectric effect. The strain/stress gradient over nanoscale thin film induces a giant flexoelectric field which act as an additional electric field besides the electrostatic field and applied field. Figure 6 (a) illustrate the 180°switched polarization state under 2.0μN tip load in 15nm PZT thin film, which is otherwise not found in cases considering pure piezoelectric effect (see II.b) and surface chemical effect (II.c). Figure 6 (b) and (c) illustrate the local distributions of out-of-plane electric field and flexoelectric field respectively. It is seen that the strain gradient induced flexoelectric field was along $[00\bar{1}]$ direction, and reached ~-2.0 MV/cm that is comparable to the electric field. Therefore the 180° switching can be attributed to this additional flexoelectric field. It should be noted this flexoelectric field increases linearly with stress gradient based on Eq. (11), thus under large tip pressure the flexoelectric effect is expected to increase. However large tip pressure will inhibit 180°switching through ferroelastic effect as illustrated in II. b. Therefore, there also exists a competing mechanism between flexoelectric and ferroelastic effect, similar to that between surface chemical and ferroelastic effect as mentioned in II. c. Figure S3 illustrates the final polarization states for film thickness/tip load combinations. Only three distinct 180° switching cases (10nm/1μN, 15nm/2μN and 20nm/4μN) were seen and highlighted by the blue ellipse (Figure S3). In the lower/left region of the ellipse, the tip pressure and thus the flexoelectric field is too small to induce 180° switching. In the upper/right region of the circle, the ferroelastic effect that favors in-plane switching overwhelms the flexoelectric effect.

Importantly, the flexoelectric effect has nontrivial dependence on the film thickness. In ultrathin PZT film (5.0nm), the stress under the tip is almost homogeneous (d$\sigma$~0) so that the stress gradient is limited, while in thicker film the stress gradient is also inhibited due to larger thickness. Therefore 180° switching is never seen in PZT thin film below 5nm or above 20nm. Consequently there is only a narrow window of thickness/pressure combination for flexoelectricity facilitated 180° switching, due to the complexity of aforementioned competing mechanism and its sensitivity on film thickness. This is best illustrated by the 1D plot of the $P_z$ along z direction beneath the tip in 10nm, 15nm and 20nm thick thin films under different tip loads [Figure 6 (d - f)], where $P_z$ near the bottom of the film ($z = 0$) decreased initially and then increased again with increasing tip loads.

## II.e. Bulk electrochemistry

Oxygen vacancies are one of the most important charged defects that are ubiquitous in many ferroelectric oxide thin films. Compared to surface charges which reside on the film surface, oxygen vacancies can diffuse into the film bulk. The redistribution of positively charged oxygen vacancies inside the film create local charges and elastic strain, which couple to the ferroelectric polarization through the electrostatic effect and Vegard strain effect. Therefore, it is important to understand the effect of oxygen vacancies on the mechanical switching behavior.

We assumed the oxygen vacancy concentration is high, $10^{21}cm^{-3}$, in PZT thin film. They were homogeneously distributed inside the film in the absence of mechanical tip load, and initially balanced by monovalent acceptors introduced from the substitution of +3 elements such as Fe on the Ti sites ($Fe_{Ti}'=A'$). The acceptors were considered fully ionized and immobile, while oxygen vacancies are mobile under external electrical/mechanical stimuli, as described from Eq. (16). For simplicity, we do not consider electronic charge effect in the current study. Figure 7 illustrates the equilibrium $P_z$ profiles in PZT thin film of different thickness subjected to different mechanical loads. In 5nm thick film, the polarization was slightly suppressed under 0.5μN mechanical load (Figure 7(a)) and switched to in-plane direction ($P_z \approx 0$) under 4μN load (Figure 7(b)). No 180° switching was seen (similar to Figure 4 (a - b)) indicating that the piezoelectric effect is dominant. In 30nm thick film under 4μN tip load, 180° switching occurred at both the tip edges on top of the film and at the bottom surface beneath the tip (Figure 7(c)). A close examination of the oxygen vacancy profile in 30nm (Figure 7(d)) indicated that oxygen vacancies migrate in both lateral direction (near the film surface) and vertical direction (towards the film bottom) away from the tip, resulting in a vacancy depleted region under the tip due to the tip induced compressive strains that disfavor the vacancies. By comparing Figure 7 (c) and (d), it is implied that the vacancy induced strain effect could potentially facilitate the switching process. Figure 7(e - f) illustrate the 1D profile of oxygen vacancies and $P_z$ along $z$ direction under the tip center in 5nm, 10nm, 20nm and 30nm thick film. It is seen that the degree of oxygen vacancy accumulation/depletion at the bottom/top surfaces increases with film thickness. This further validates why 180° switching is only seen in thicker films (>20nm), as vacancy transport and the Vegard strain effect is too small to facilitate the switching in films less than 10nm thick.

**II.f. Pressure-thickness phase-diagram of mechanical switching**

We thus calculated the final polarization state under different film thickness and tip load conditions, by turning on only one effect (surface charge, flexoelectricity or bulk oxygen vacancy) at a time while freezing the others (Figure S1 - S4). Based on this we constructed four separate pressure-thickness phase diagrams for each effect, as illustrated in Figure 8 (a - d). Pure piezoelectric effect is insufficient for 180° switching under any tip loads in PZT thin film of any thickness (Figure 8(a)), while the presence of surface charges and oxygen vacancies could enable 180° switching in ultrathin film (<10nm, Figure 8(b)) and thicker film (>20nm, Figure 8(d)) respectively. The flexoelectric effect could potentially result in 180° switching in intermediate range of film thickness (10~20nm) depending on certain tip pressures (Figure 8(c)). We thus combined Figure 8 (a - d) into a general phase diagram of 180° switchable/non-switchable PZT thin film of different thickness under different tip loads, as illustrated in Figure 9. In ultrathin film (<5nm), the 180° switching is driven by the surface chemical effect (circles in red region in Figure 9), while in thicker film (>20nm) the bulk oxygen vacancy induced Vegard strain effect is dominant (squares in blue region in Figure 9). The surface charge facilitated switching occurred at any tip loads in 5nm thick film; however it only occurred under 1μN load in 10nm thick film.

This is due to the competition between the electrochemical potential that promotes the 180° switching and piezoelectric effect that favors the in-plane orientation, both of which scale with the tip pressure. On the other hand, the bulk vacancy facilitated switching occurred in thicker film under large tip pressure. This is because significant oxygen vacancy migration and local segregation requires large tip pressure and enough diffusion length. Finally, the flexoelectric effect becomes dominant in thin films from 10~20nm thick under certain tip loads in a narrow "channel" (triangles in green region in Figure 9). The critical switching pressure increases with film thickness. This is due to the fact that the flexoelectric field depends on strain gradient which scales with both tip load and film thickness. In ultrathin film (<5nm), the stress across the film becomes homogeneous that lacks stress gradient, while in thicker film (>20nm) the strain gradient is localized only near the film surface under the tip and is insufficient to switch the entire film.

**III. Coupled modelling**

One of the advantages of phase-field simulation is that we can not only study the response of certain properties (such as mechanical switching) to one particular effect independently by freezing other effects, but also tune the strength of this effect to study the linearity of its response, and turn on more effects to understand their coupling behaviors. As an example we turned on all the aforementioned effects (surface charge, flexoelectricity and bulk vacancy dynamics) to study the mechanical switching behavior in a 20nm PZT thin film subjected to 4μN tip load. Figure 10 illustrates the magnitude of $P_z$ on top surface under the tip ($P_{z\_top}$) and on bottom surface beneath the tip ($P_{z\_bottom}$), at different oxygen vacancy concentration $V_O^{\bullet\bullet}$ and longitudinal flexoelectric strength ($F_{11}$). It is seen that $P_{z\_top}$ decreases exponentially with increasing vacancy concentration at given $F_{11}$ [Figure 10(a)], while it decreases almost linearly with increasing $F_{11}$ at fixed vacancy concentration [Figure 10(b)]. This indicated that the flexoelectric field effect on $P_{z\_top}$ is instantaneous and linear, while the Vegard strain effect on $P_{z\_top}$ only becomes significant when vacancy concentration is above a threshold. On the other hand, sudden jumps of $P_{z\_bottom}$ from positive to negative polarity were seen at critical vacancy concentrations [Figure 10(c)]. This critical concentration is independent of $F_{11}$ when $F_{11}$ is smaller than 2.0 ($10^{-11}Vm^3N^{-1}$), indicating that flexoelectric effect is negligible in this region. The critical concentration becomes smaller (~$10^{20}cm^{-3}$) when $F_{11}$ reaches 5.0 ($10^{-11}Vm^3N^{-1}$) and eventually disappear when $F_{11}$ is 10 ($10^{-11}Vm^3N^{-1}$), implying that flexoelectric field becomes dominant. Similarly, the critical flexoelectric strength for $P_{z\_bottom}$ switching becomes smaller with increasing oxygen vacancy concentration and finally $P_{z\_bottom}$ becomes switchable at any $F_{11}$ [Figure 10(d)]. Analysis on the coupled flexoelectric and Vegard strain effect allows us to understand which effect is dominant under what condition.

It should be noted that in this study we do not consider the kinetic process of each effect, i.e., all the effects are assumed to reach steady state with final ferroelectric polarization state. In fact, ferroelectric switching under mechanical pressure can be very fast and the flexoelectric effect occurs instantaneously in the presence of strain gradient. On the other hand, surface electrochemistry is limited by the transport rate of ionic species along the surfaces (available from time resolved PFM data [85,86]) and by transport from gas phase to the film surface. And the Vegard strain effect is limited by the oxygen vacancy diffusivity, which is normally very slow in solids under room temperature, but becomes much faster once the local electric bias and mechanical load overcome the activation energy barrier of vacancies. These makes the mechanical switching dynamics more complicated and is not taken into account in the current model. Nevertheless, our model still clarifies some important issues about the recently reported pressure induced writing, and provide a clear picture about the possibility of multi-effect in pressure switching and in which realm (spatial and pressure combination) they become dominant.

## IV. Summary

We developed a phase-field model to explore the surface electrochemical phenomena, bulk flexoelectric effect and bulk vacancy dynamics as thermodynamic driving force to the pressure induced mechanical switching in $Pb(Zr_{0.2}Ti_{0.8})O_3$ ferroelectric thin film. We found that piezoelectric effect from pure mechanical pressure can only rotate the out-of-plane polarization to the in-plane orientation and no 180° switching is observed. Surface charges shift the electrochemical potential that is responsible for the 180° switching in ultrathin film, while bulk Vegard strain effect is dominant and accounts for the 180° switching in thicker film under large tip pressure. The flexoelectricity induced switching becomes more important in thin film at intermediate thickness (10~20nm). A linearity analysis on polarization field dependence on vacancy concentration and flexoelectric strength indicate that all three effects couple and interact with each other. In this study, we only focus on the final steady state of polarization, surface charge and bulk vacancy and do not consider the kinetics process of each of them. Our work successfully differential the surface phenomena, long-range strain gradient and bulk vacancy transport dynamics in the mechanical switching behavior, providing a clear picture of thickness /pressure dependence of mechanical switching behavior in ferroelectric thin film.

**Acknowledgements**

This study was supported by the U.S. DOE, Office of Basic Energy Sciences (BES), Materials Sciences and Engineering Division (MSED) under FWP Grant No. ERKCZ07 (Y.C., S.V.K.). A portion of this research was conducted at the Center for Nanophase Materials Sciences, which is a DOE Office of Science User Facility.

**Figures:**

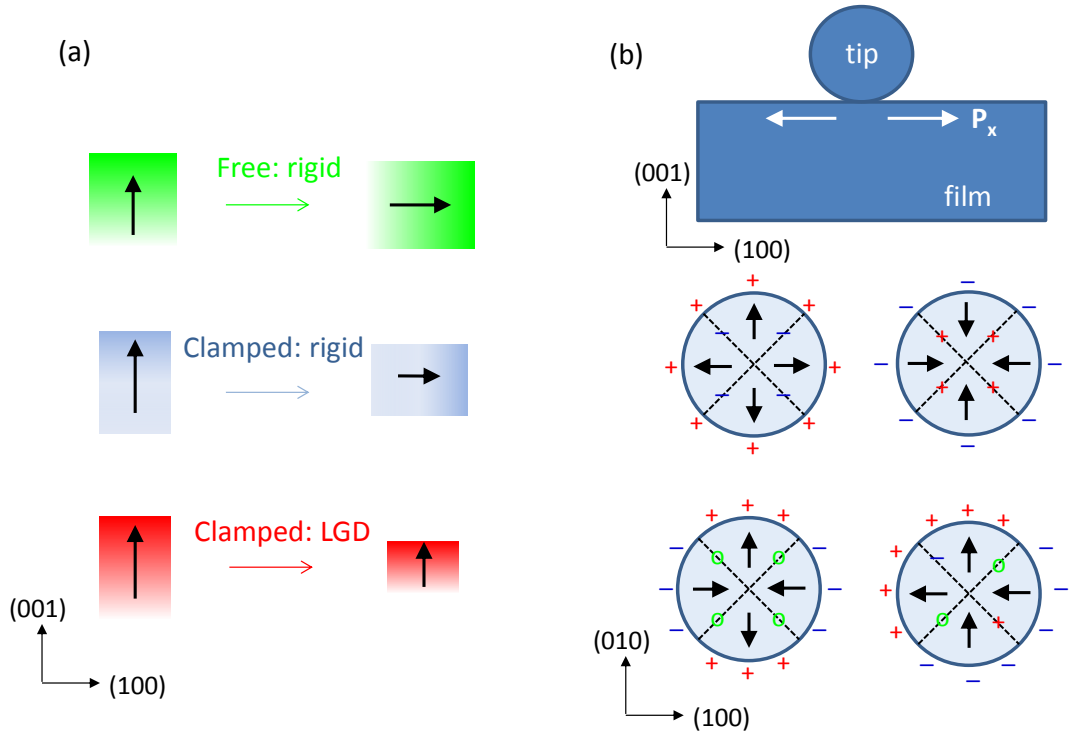

Figure 1, Schematic illustration of polarization suppressions and rotations from (001) to (100) orientations under applied pressure for unclamped crystal and clamped crystals (a), and some possible domain configurations of different in-plane polarization components creating positively (+), negatively (-) charged and neutral (o) domain walls under the tip (b).

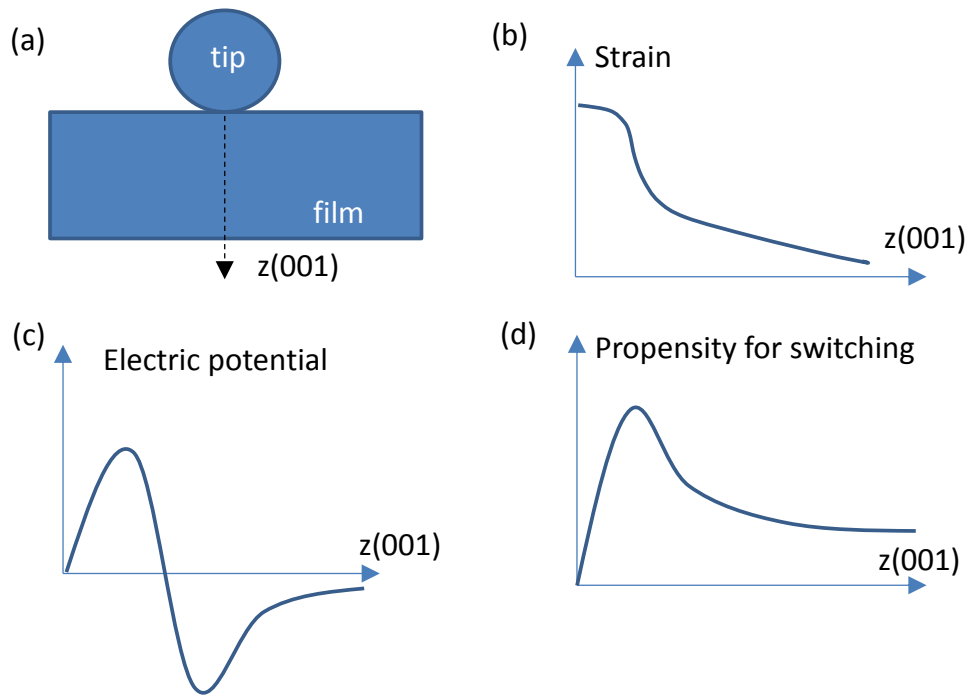

Figure 2, Schematic 1D profiles of elastic strain (b), electric potential (c) and propensity for switching (d) along (001) direction in ferroelectric thin film subjected to tip pressure in SPM geometry (a).

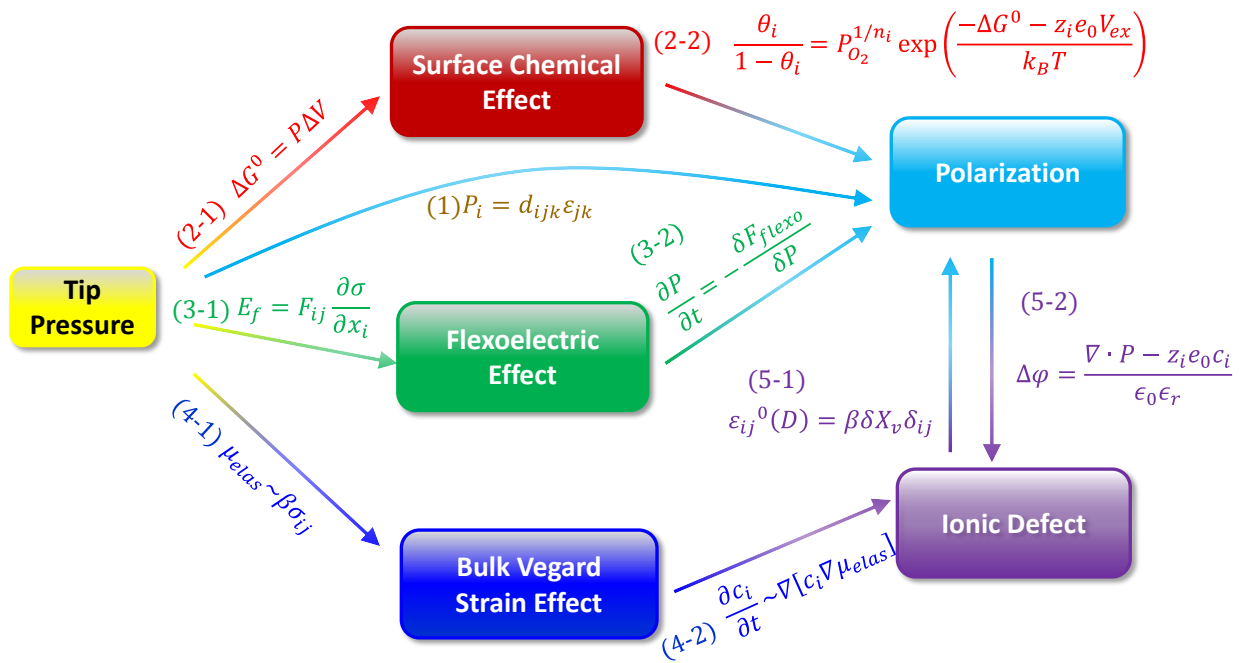

Figure 3, Schematic illustration and equations of multi-effect coupling including (1) pure piezoelectricity; (2) surface chemistry; (3) flexoelectricity and (4) bulk Vegard strain effect, and their effects on ferroelectric polarization and ionic defects (5)

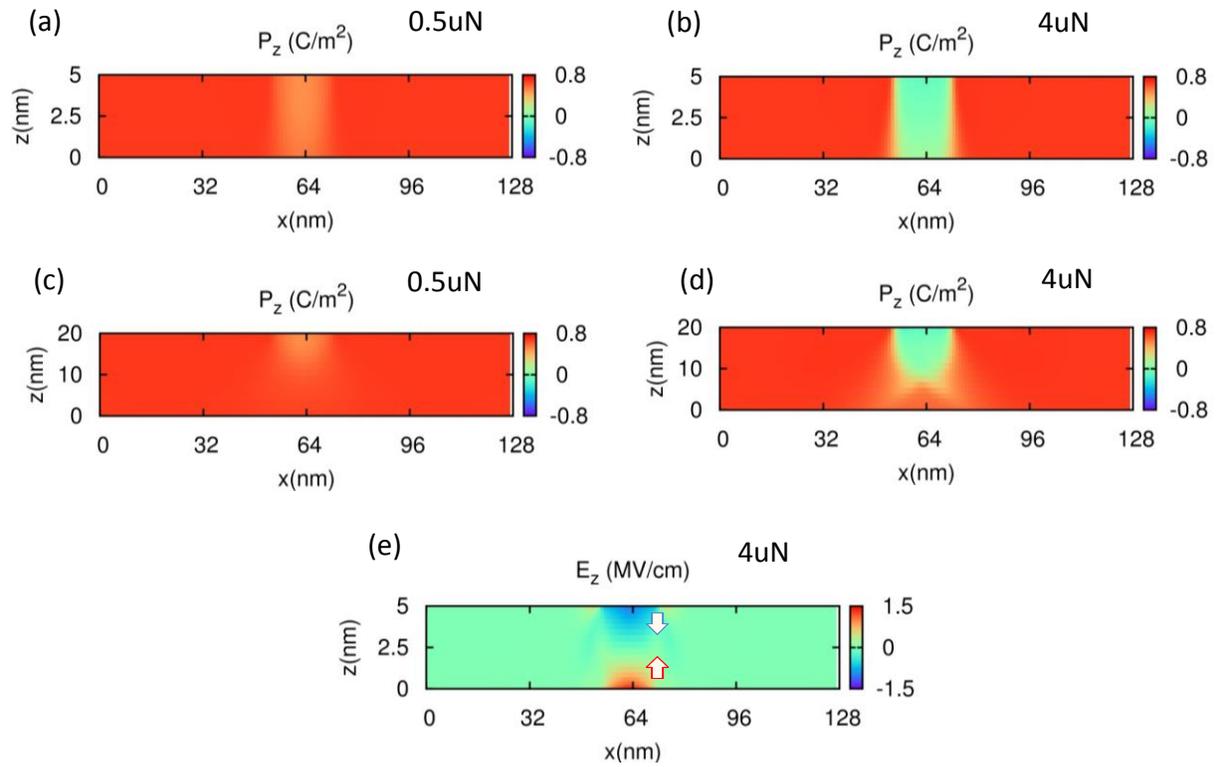

Figure 4, Vertical polarization ($P_z$) profiles in 5nm thick (a - b) and 20nm thick (c - d) PZT thin film with pure piezoelectric effect. The applied tip load is 0.5μN (a), (c) and 4μN (b), (d) respectively. (e) The out-of-plane electric field distribution ($E_z$) in 5nm thin film under 4μN

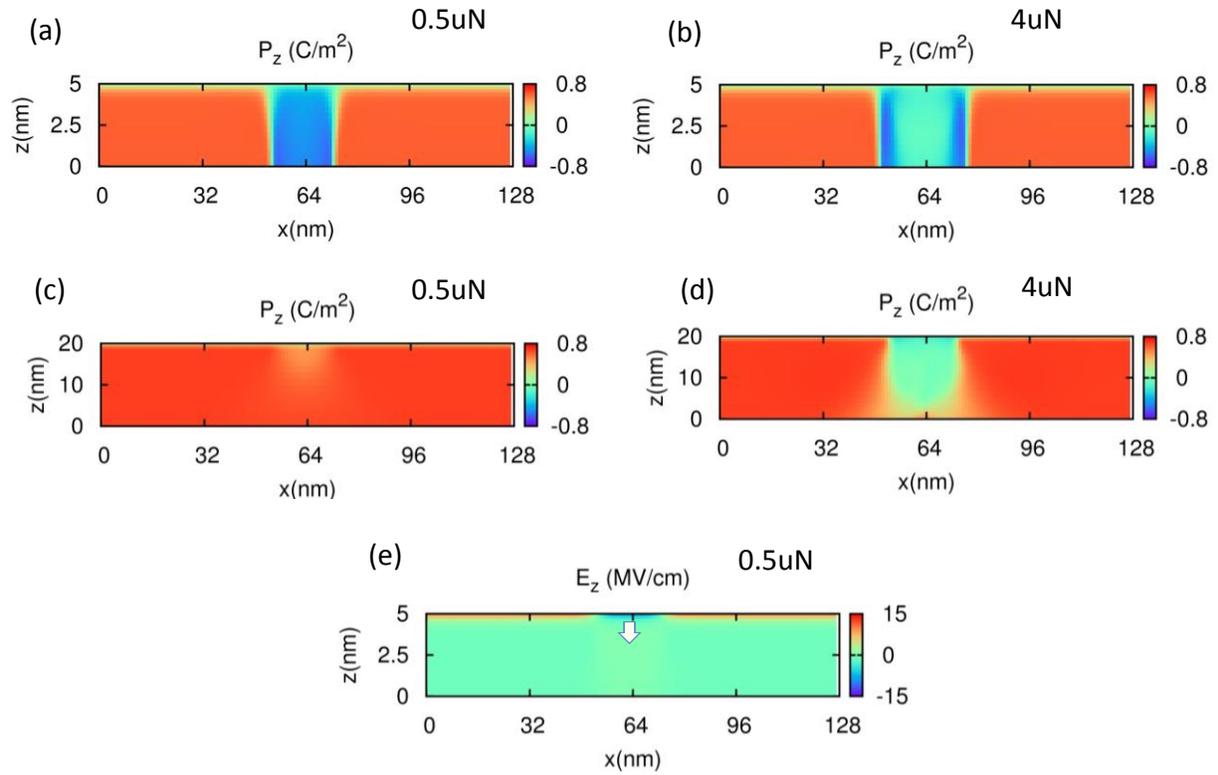

Figure 5, $P_z$ profiles in 5nm thick (a - b) and 20nm thick (c - d) PZT thin film with surface electrochemical effect. The applied tip load is 0.5μN (a), (c) and 4μN (b), (d) respectively. (e) The out-of-plane electric field distribution ($E_z$) in 5nm thin film under 0.5μN

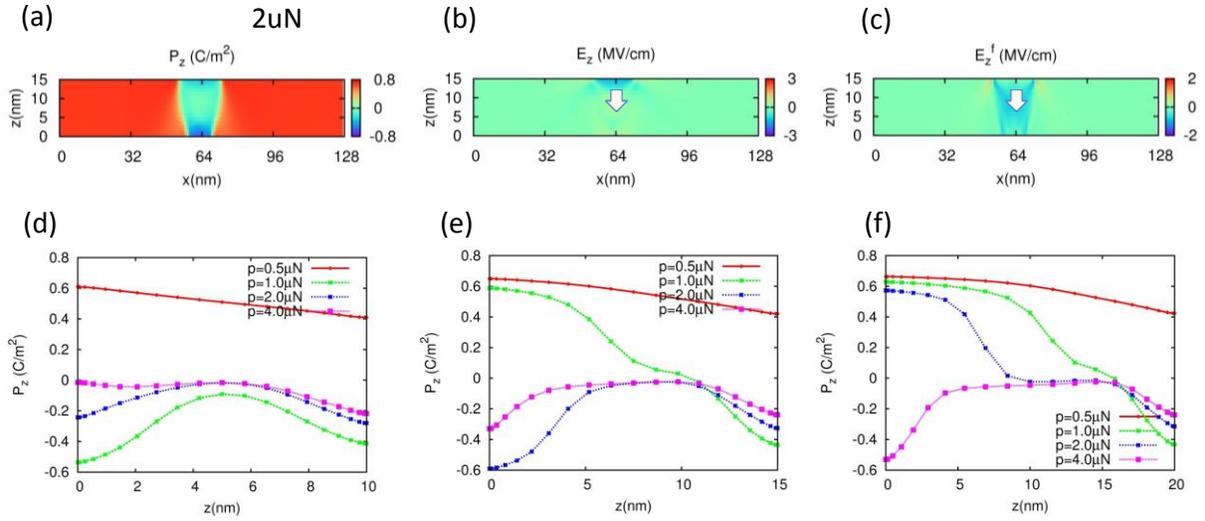

Figure 6, $P_z$ profiles (a), vertical electrostatic field ($E_z$) (b), and vertical flexoelectric field ($E_z^f$) (c) in 15nm thick PZT thin film under 2μN tip load with flexoelectric effect. 1D profiles of $P_z$ along z direction beneath the tip in 10nm thick (d), 15nm thick (e), and 20nm thick (f) PZT thin film under different tip loads

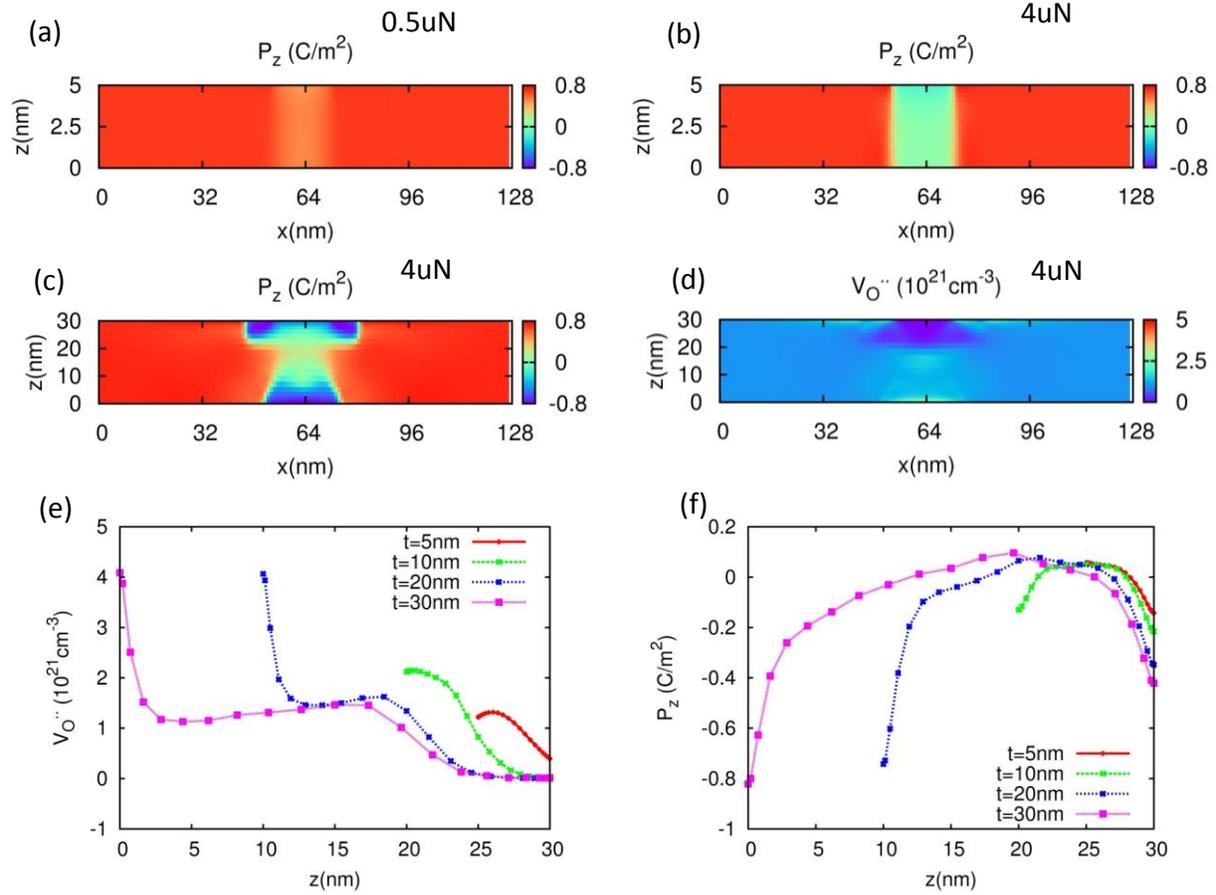

Figure 7, $P_z$ profiles in 5nm thick PZT film under 0.5μN (a) and 4μN (b) tip loads, and $P_z$ profile (c) and oxygen vacancy concentration (d) in 30nm thick PZT film under 4μN tip load. 1D profiles of (e) oxygen vacancy concentration and (f) $P_z$ along z direction beneath the tip in PZT thin films of different thickness (5, 10, 20 and 30nm) under 4μN tip load (initial homogeneous oxygen vacancy concentration is $10^{21}$ cm$^{-3}$).

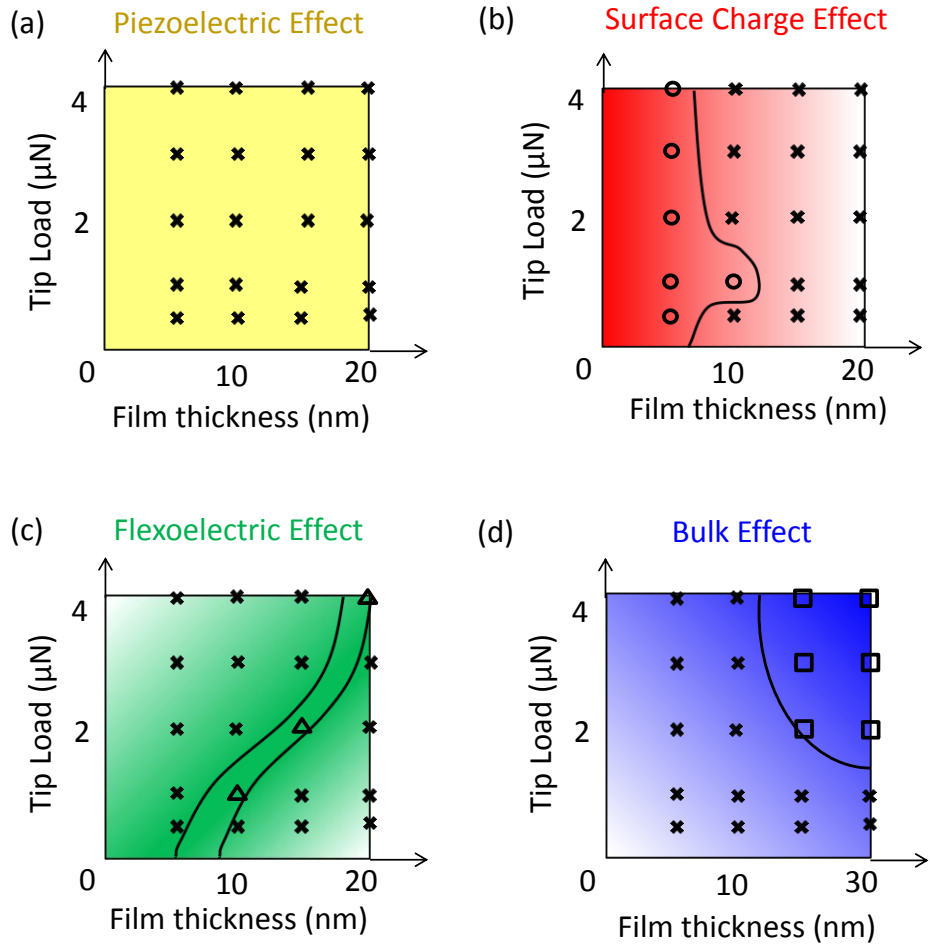

Figure 8 Pressure-film thickness phase diagrams of 180 ° mechanical switching for each separate effect including only (a) pure ferroelastic effect; (b) ferroelastic and surface charge effect; (c) ferroelastic and flexoelectric effect and (d) ferroelastic and bulk chemical effect.

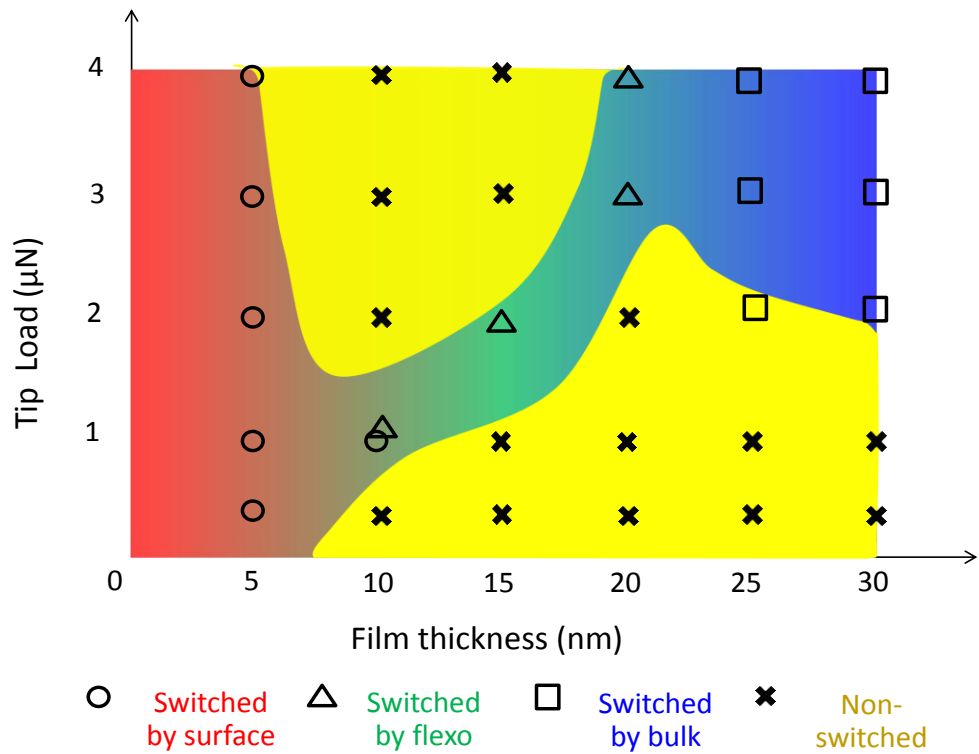

Figure 9, Pressure-thickness phase-diagram of 180 ° mechanical switching with multi-effect. The red region with circles indicates that the switching is facilitated by surface charge effect. The green region with triangles indicates that the switching is facilitated by flexoelectric effect. And the blue region with squares indicates that the switching is facilitated by bulk Vegard strain effect. The yellow region with cross is non-switchable region.

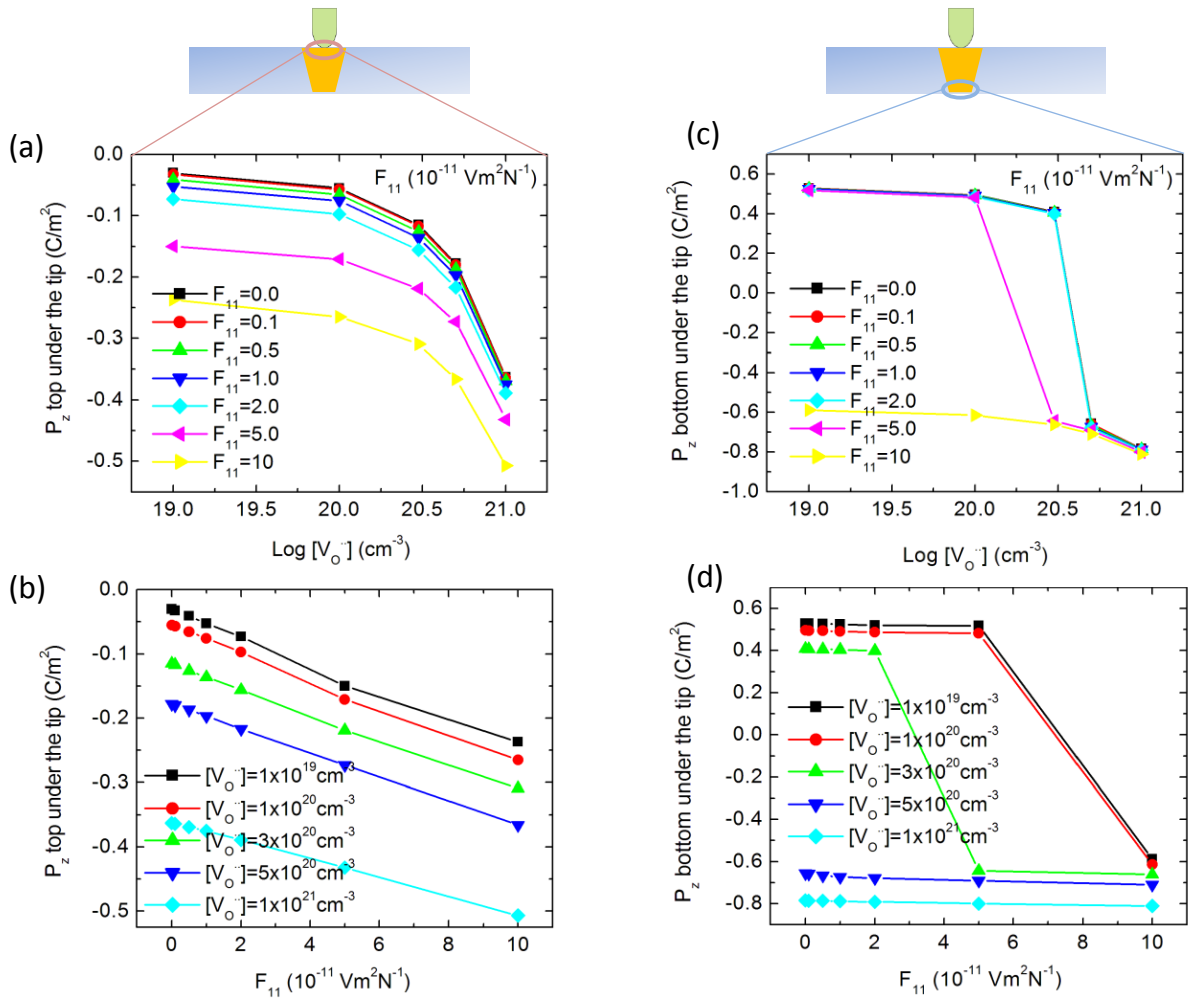

Figure 10, Dependence of $P_z$ on top surface under the tip on the oxygen vacancy concentration (a) and longitudinal flexoelectric strength (b), and $P_z$ on bottom surface below the tip on the oxygen vacancy concentration (c) and longitudinal flexoelectric strength (d)

# References

1. Maksymovych, P. et al. Dynamic Conductivity of Ferroelectric Domain Walls in BiFeO(3). *Nano Lett* **11**, 1906-1912, doi:10.1021/nl104363x (2011).
2. Seidel, J. et al. Domain Wall Conductivity in La-Doped BiFeO(3). *Phys Rev Lett* **105**, doi:197603

10.1103/PhysRevLett.105.197603 (2010).
3. Seidel, J. et al. Conduction at domain walls in oxide multiferroics. *Nat Mater* **8**, 229-234, doi:10.1038/nmat2373 (2009).
4. Gruverman, A. et al. Tunneling Electroresistance Effect in Ferroelectric Tunnel Junctions at the Nanoscale. *Nano Lett* **9**, 3539-3543, doi:10.1021/nl901754t (2009).
5. Tsymbal, E. Y. & Kohlstedt, H. Applied physics - Tunneling across a ferroelectric. *Science* **313**, 181-183, doi:10.1126/science.1126230 (2006).
6. Bocher, L. et al. Atomic and Electronic Structure of the BaTiO3/Fe Interface in Multiferroic Tunnel Junctions. *Nano Lett* **12**, 376-382, doi:10.1021/nl203657c (2012).
7. Scott, J. F. Applications of modern ferroelectrics. *Science* **315**, 954-959, doi:10.1126/science.1129564 (2007).
8. Gruverman, A., Auciello, O., Ramesh, R. & Tokumoto, H. Scanning force microscopy of domain structure in ferroelectric thin films: imaging and control. *Nanotechnology* **8**, A38-A43, doi:10.1088/0957-4484/8/3a/008 (1997).
9. Gruverman, A. L., Hatano, J. & Tokumoto, H. Scanning force microscopy studies of domain structure in BaTiO3 single crystals. *Japanese Journal of Applied Physics Part 1-Regular Papers Short Notes & Review Papers* **36**, 2207-2211, doi:10.1143/jjap.36.2207 (1997).
10. Balke, N., Bdikin, I., Kalinin, S. V. & Kholkin, A. L. Electromechanical Imaging and Spectroscopy of Ferroelectric and Piezoelectric Materials: State of the Art and Prospects for the Future. *J Am Ceram Soc* **92**, 1629-1647, doi:10.1111/j.1551-2916.2009.03240.x (2009).
11. Bonnell, D. A., Kalinin, S. V., Kholkin, A. L. & Gruverman, A. Piezoresponse Force Microscopy: A Window into Electromechanical Behavior at the Nanoscale. *Mrs Bulletin* **34**, 648-657, doi:10.1557/mrs2009.176 (2009).
12. Kalinin, S. V., Setter, N. & Kholkin, A. L. Electromechanics on the Nanometer Scale: Emerging Phenomena, Devices, and Applications. *Mrs Bulletin* **34**, 634-642, doi:10.1557/mrs2009.174 (2009).
13. Gruverman, A. & Kalinin, S. V. Piezoresponse force microscopy and recent advances in nanoscale studies of ferroelectrics. *Journal of Materials Science* **41**, 107-116, doi:10.1007/s10853-005-5946-0 (2006).
14. Chen, L. et al. Formation of 90 degrees elastic domains during local 180 degrees switching in epitaxial ferroelectric thin films. *Applied Physics Letters* **84**, 254-256, doi:10.1063/1.1633970 (2004).
15. Ganpule, C. S. et al. Polarization relaxation kinetics and 180 degrees domain wall dynamics in ferroelectric thin films. *Phys Rev B* **65**, doi:014101

10.1103/PhysRevB.65.014101 (2002).
16. Roelofs, A. et al. Depolarizing-field-mediated 180 degrees switching in ferroelectric thin films with 90 degrees domains. *Applied Physics Letters* **80**, 1424-1426, doi:10.1063/1.1448653 (2002).
17. Paruch, P., Giamarchi, T. & Triscone, J. M. in *Physics of Ferroelectrics: A Modern Perspective* Vol. 105 *Topics in Applied Physics* (eds K. M. Rabe, C. H. Ahn, & J. M. Triscone) 339-362 (Springer-Verlag Berlin, 2007).
18. Paruch, P., Giamarchi, T. & Triscone, J. M. Domain wall roughness in epitaxial ferroelectric PbZr(0.2)Ti(0.8)O(3) thin films. *Phys Rev Lett* **94**, doi:197601

10.1103/PhysRevLett.94.197601 (2005).
19    Tybell, T., Paruch, P., Giamarchi, T. & Triscone, J. M. Domain wall creep in epitaxial ferroelectric Pb(Zr(0.2)Ti(0.8))O(3) thin films. *Phys Rev Lett* **89**, doi:097601

10.1103/PhysRevLett.89.097601 (2002).
20    Aravind, V. R. *et al.* Correlated polarization switching in the proximity of a 180 degrees domain wall. *Phys Rev B* **82**, doi:024111

10.1103/PhysRevB.82.024111 (2010).
21    Kalinin, S. V. *et al.* Spatial distribution of relaxation behavior on the surface of a ferroelectric relaxor in the ergodic phase. *Applied Physics Letters* **95**, 142902, doi:10.1063/1.3242011 (2009).
22    Kalinin, S. V. *et al.* Imaging mechanism of piezoresponse force microscopy in capacitor structures. *Applied Physics Letters* **92**, 152906, doi:10.1063/1.2905266 (2008).
23    Jesse, S. *et al.* Direct imaging of the spatial and energy distribution of nucleation centres in ferroelectric materials. *Nat Mater* **7**, 209-215, doi:10.1038/nmat2114 (2008).
24    Kalinin, S. V., Karapetian, E. & Kachanov, M. Nanoelectromechanics of piezoresponse force microscopy. *Phys Rev B* **70**, 184101, doi:10.1103/PhysRevB.70.184101 (2004).
25    Kalinin, S. V., Shin, J., Kachanov, M., Karapetian, E. & Baddorf, A. P. in *Ferroelectric Thin Films Xii* Vol. 784 *Materials Research Society Symposium Proceedings* (eds S. HoffmannEifert *et al.*)  43-48 (2004).
26    Kalinin, S. V., Eliseev, E. A. & Morozovska, A. N. Materials contrast in piezoresponse force microscopy. *Applied Physics Letters* **88**, doi:232904

10.1063/1.2206992 (2006).
27    Morozovska, A. N., Eliseev, E. A. & Kalinin, S. V. Domain nucleation and hysteresis loop shape in piezoresponse force spectroscopy. *Applied Physics Letters* **89**, doi:192901

10.1063/1.2378526 (2006).
28    Morozovska, A. N., Eliseev, E. A., Bravina, S. L. & Kalinin, S. V. Resolution-function theory in piezoresponse force microscopy: Wall imaging, spectroscopy, and lateral resolution. *Phys Rev B* **75**, 174109, doi:10.1103/PhysRevB.75.174109 (2007).
29    Morozovska, A. N., Eliseev, E. A. & Kalinin, S. V. The piezoresponse force microscopy of surface layers and thin films: Effective response and resolution function. *J Appl Phys* **102**, doi:074105

10.1063/1.2785824 (2007).
30    Morozovska, A. N. *et al.* Local polarization switching in the presence of surface-charged defects: Microscopic mechanisms and piezoresponse force spectroscopy observations. *Phys Rev B* **78**, doi:054101

10.1103/PhysRevB.78.054101 (2008).
31    Kalinin, S. V. *et al.* Nanoelectromechanics of polarization switching in piezoresponse force microscopy. *J Appl Phys* **97**, doi:074305

10.1063/1.1866483 (2005).
32    Molotskii, M. I. & Shvebelman, M. M. Dynamics of ferroelectric domain formation in an atomic force microscope. *Philosophical Magazine* **85**, 1637-1655, doi:10.1080/14786430312331524670 (2005).
33    Molotskii, M. & Winebrand, E. Interactions of an atomic force microscope tip with a reversed ferroelectric domain. *Phys Rev B* **71**, doi:132103

10.1103/PhysRevB.71.132103 (2005).
34   Rosenman, G., Shur, D., Garb, K., Cohen, R. & Krasik, Y. E. Polarization switching in ferroelectric cathodes. *J Appl Phys* **82**, 772-778, doi:Doi 10.1063/1.365771 (1997).
35   Morozovska, A. N., Eliseev, E. A., Bravina, S. L. & Kalinin, S. V. Landau-Ginzburg-Devonshire theory for electromechanical hysteresis loop formation in piezoresponse force microscopy of thin films. *J Appl Phys* **110**, doi:052011

10.1063/1.3623763 (2011).
36   Morozovska, A. N. *et al.* Thermodynamics of nanodomain formation and breakdown in scanning probe microscopy: Landau-Ginzburg-Devonshire approach. *Phys Rev B* **80**, doi:214110

10.1103/PhysRevB.80.214110 (2009).
37   Kalinin, S. V. *et al.* Intrinsic single-domain switching in ferroelectric materials on a nearly ideal surface. *Proceedings of the National Academy of Sciences of the United States of America* **104**, 20204-20209, doi:10.1073/pnas.0709316104 (2007).
38   Kalinin, S. V., Morozovska, A. N., Chen, L. Q. & Rodriguez, B. J. Local polarization dynamics in ferroelectric materials. *Reports on Progress in Physics* **73**, 056502, doi:10.1088/0034-4885/73/5/056502 (2010).
39   Balke, N. *et al.* Deterministic control of ferroelastic switching in multiferroic materials. *Nature Nanotechnology* **4**, 868-875, doi:10.1038/nnano.2009.293 (2009).
40   Xiao, Y., Shenoy, V. B. & Bhattacharya, K. Depletion layers and domain walls in semiconducting ferroelectric thin films. *Phys Rev Lett* **95**, doi:Artn 247603

10.1103/Physrevlett.95.247603 (2005).
41   Xiao, Y. & Bhattacharya, K. A continuum theory of deformable, semiconducting ferroelectrics. *Archive for Rational Mechanics and Analysis* **189**, 59-95, doi:10.1007/s00205-007-0096-y (2008).
42   Lu, H. D. *et al.* Nanodomain Engineering in Ferroelectric Capacitors with Graphene Electrodes. *Nano Lett* **16**, 6460-6466, doi:10.1021/acs.nanolett.6b02963 (2016).
43   Lu, H. *et al.* Mechanical Writing of Ferroelectric Polarization. *Science* **336**, 59-61, doi:10.1126/science.1218693 (2012).
44   Sharma, P. *et al.* Mechanical Tuning of LaAlO3/SrTiO3 Interface Conductivity. *Nano Lett* **15**, 3547-3551, doi:10.1021/acs.nanolett.5b01021 (2015).
45   Ocenasek, J. *et al.* Nanomechanics of flexoelectric switching. *Phys Rev B* **92**, doi:Artn 035417

10.1103/Physrevb.92.035417 (2015).
46   Borisevich, A. Y. *et al.* Interface dipole between two metallic oxides caused by localized oxygen vacancies. *Phys Rev B* **86**, doi:140102

10.1103/PhysRevB.86.140102 (2012).
47   Kalinin, S. V., Borisevich, A. & Fong, D. Beyond Condensed Matter Physics on the Nanoscale: The Role of Ionic and Electrochemical Phenomena in the Physical Functionalities of Oxide Materials. *Acs Nano* **6**, 10423-10437, doi:10.1021/nn304930x (2012).
48   Kalinin, S. V. & Spaldin, N. A. Functional Ion Defects in Transition Metal Oxides. *Science* **341**, 858-859, doi:10.1126/science.1243098 (2013).
49   Kalinin, S. V., Jesse, S., Tselev, A., Baddorf, A. P. & Balke, N. The Role of Electrochemical Phenomena in Scanning Probe Microscopy of Ferroelectric Thin Films. *Acs Nano* **5**, 5683-5691, doi:10.1021/nn2013518 (2011).

50  Ievlev, A. V. *et al.* Chemical State Evolution in Ferroelectric Films during Tip-Induced Polarization and Electroresistive Switching. *Acs Applied Materials & Interfaces* **8**, 29588-29593, doi:10.1021/acsami.6b10784 (2016).
51  Jesse, S. *et al.* Direct Mapping of Ionic Transport in a Si Anode on the Nanoscale: Time Domain Electrochemical Strain Spectroscopy Study. *Acs Nano* **5**, 9682-9695, doi:10.1021/nn203141g (2011).
52  Kumar, A., Ciucci, F., Morozovska, A. N., Kalinin, S. V. & Jesse, S. Measuring oxygen reduction/evolution reactions on the nanoscale. *Nature Chemistry* **3**, 707-713, doi:10.1038/nchem.1112 (2011).
53  Morozovska, A. N., Eliseev, E. A., Svechnikov, G. S. & Kalinin, S. V. Nanoscale electromechanics of paraelectric materials with mobile charges: Size effects and nonlinearity of electromechanical response of SrTiO(3) films. *Phys Rev B* **84**, 045402, doi:10.1103/PhysRevB.84.045402 (2011).
54  Morozovska, A. N. *et al.* Thermodynamics of electromechanically coupled mixed ionic-electronic conductors: Deformation potential, Vegard strains, and flexoelectric effect. *Phys Rev B* **83**, doi:195313

10.1103/PhysRevB.83.195313 (2011).
55  Morozovska, A. N., Eliseev, E. A., Balke, N. & Kalinin, S. V. Local probing of ionic diffusion by electrochemical strain microscopy: Spatial resolution and signal formation mechanisms. *J Appl Phys* **108**, 053712, doi:10.1063/1.3460637 (2010).
56  Morozovska, A. N., Eliseev, E. A. & Kalinin, S. V. Electromechanical probing of ionic currents in energy storage materials. *Applied Physics Letters* **96**, doi:222906

10.1063/1.3446838 (2010).
57  Balke, N. *et al.* Nanoscale mapping of ion diffusion in a lithium-ion battery cathode. *Nature Nanotechnology* **5**, 749-754, doi:10.1038/nnano.2010.174 (2010).
58  Gureev, M. Y., Tagantsev, A. K. & Setter, N. Head-to-head and tail-to-tail 180 degrees domain walls in an isolated ferroelectric. *Phys Rev B* **83**, doi:184104

10.1103/PhysRevB.83.184104 (2011).
59  Morozovska, A. N. *et al.* Flexocoupling impact on size effects of piezoresponse and conductance in mixed-type ferroelectric semiconductors under applied pressure. *Phys Rev B* **94**, doi:Artn 174101

10.1103/Physrevb.94.174101 (2016).
60  Haun, M. J., Zhuang, Z. Q., Furman, E., Jang, S. J. & Cross, L. E. THERMODYNAMIC THEORY OF THE LEAD ZIRCONATE-TITANATE SOLID-SOLUTION SYSTEM, .3. CURIE CONSTANT AND 6TH-ORDER POLARIZATION INTERACTION DIELECTRIC STIFFNESS COEFFICIENTS. *Ferroelectrics* **99**, 45-54, doi:10.1080/00150198908221438 (1989).
61  Pertsev, N. A., Kukhar, V. G., Kohlstedt, H. & Waser, R. Phase diagrams and physical properties of single-domain epitaxial Pb(Zr1-xTix)O-3 thin films. *Phys Rev B* **67**, doi:Artn 054107

10.1103/Physrevb.67.054107 (2003).
62  Freedman, D. A., Roundy, D. & Arias, T. A. Elastic effects of vacancies in strontium titanate: Short- and long-range strain fields, elastic dipole tensors, and chemical strain. *Phys Rev B* **80**, doi:064108

10.1103/PhysRevB.80.064108 (2009).

63  Kretschmer, R. & Binder, K. SURFACE EFFECTS ON PHASE-TRANSITIONS IN FERROELECTRICS AND DIPOLAR MAGNETS. *Phys Rev B* **20**, 1065-1076, doi:10.1103/PhysRevB.20.1065 (1979).
64  Jia, C. L. *et al.* Unit-cell scale mapping of ferroelectricity and tetragonality in epitaxial ultrathin ferroelectric films. *Nat Mater* **6**, 64-69, doi:10.1038/nmat1808 (2007).
65  Wang, J., Tagantsev, A. K. & Setter, N. Size effect in ferroelectrics: Competition between geometrical and crystalline symmetries. *Phys Rev B* **83**, doi:Artn 014104

10.1103/Physrevb.83.014104 (2011).
66  Pertsev, N. A., Zembilgotov, A. G. & Tagantsev, A. K. Effect of mechanical boundary conditions on phase diagrams of epitaxial ferroelectric thin films. *Phys Rev Lett* **80**, 1988-1991, doi:DOI 10.1103/PhysRevLett.80.1988 (1998).
67  Kukhar, V. G., Pertsev, N. A., Kohlstedt, H. & Waser, R. Polarization states of polydomain epitaxial Pb(Zr(1-x)Ti(x))O(3) thin films and their dielectric properties. *Phys Rev B* **73**, doi:Artn 214103

10.1103/Physrevb.73.214103 (2006).
68  Karapetian, E., Kachanov, M. & Kalinin, S. V. Nanoelectromechanics of piezoelectric indentation and applications to scanning probe microscopies of ferroelectric materials. *Philosophical Magazine* **85**, 1017-1051, doi:10.1080/14786430412331324680 (2005).
69  Pan, K., Liu, Y. Y., Xie, S. H., Liu, Y. M. & Li, J. Y. The electromechanics of piezoresponse force microscopy for a transversely isotropic piezoelectric medium. *Acta Mater* **61**, 7020-7033, doi:10.1016/j.actamat.2013.08.019 (2013).
70  Chen, W. Q., Pan, E. N., Wang, H. M. & Zhang, C. Z. Theory of indentation on multiferroic composite materials. *Journal of the Mechanics and Physics of Solids* **58**, 1524-1551, doi:10.1016/j.jmps.2010.07.012 (2010).
71  Chen, W. Q. & Ding, H. J. Indentation of a transversely isotropic piezoelectric half-space by a rigid sphere. *Acta Mechanica Solida Sinica* **12**, 114-120 (1999).
72  Eliseev, E. A., Morozovska, A. N., Svechnikov, G. S., Maksymovych, P. & Kalinin, S. V. Domain wall conduction in multiaxial ferroelectrics. *Phys Rev B* **85**, 045312, doi:10.1103/PhysRevB.85.045312 (2012).
73  Zubko, P., Catalan, G., Buckley, A., Welche, P. R. L. & Scott, J. F. Strain-gradient-induced polarization in SrTiO3 single crystals. *Phys Rev Lett* **99**, 167601, doi:10.1103/PhysRevLett.99.167601 (2007).
74  Zubko, P., Catalan, G., Buckley, A., Welche, P. R. L. & Scott, J. F. Strain-gradient-induced polarization in SrTiO(3) single crystals (vol 99, art no 167601, 2007). *Phys Rev Lett* **100**, doi:Artn 199906

10.1103/Physrevlett.100.199906 (2008).
75  Biancoli, A., Fancher, C. M., Jones, J. L. & Damjanovic, D. Breaking of macroscopic centric symmetry in paraelectric phases of ferroelectric materials and implications for flexoelectricity. *Nat Mater* **14**, 224-229, doi:10.1038/NMAT4139 (2015).
76  Morozovska, A. N., Eliseev, E. A., Morozovsky, N. V. & Kalinin, S. V. Ferroionic states in ferroelectric thin films. *Phys Rev B* **95**, doi:Artn 195413

10.1103/Physrevb.95.195413 (2017).
77  Huang, J. S., Sumpter, B. G. & Meunier, V. A universal model for nanoporous carbon supercapacitors applicable to diverse pore regimes, carbon materials, and electrolytes. *Chem-Eur J* **14**, 6614-6626, doi:10.1002/chem.200800639 (2008).

78  Wu, P., Huang, J. S., Meunier, V., Sumpter, B. G. & Qiao, R. Complex Capacitance Scaling in Ionic Liquids-Filled Nanopores. *Acs Nano* **5**, 9044-9051, doi:10.1021/nn203260w (2011).
79  Wu, P., Huang, J. S., Meunier, V., Sumpter, B. G. & Qiao, R. Voltage Dependent Charge Storage Modes and Capacity in Subnanometer Pores. *J Phys Chem Lett* **3**, 1732-1737, doi:10.1021/jz300506j (2012).
80  Bazant, M. Z., Thornton, K. & Ajdari, A. Diffuse-charge dynamics in electrochemical systems. *Physical Review E* **70**, 021506, doi:10.1103/PhysRevE.70.021506 (2004).
81  Collins, L. *et al.* Probing charge screening dynamics and electrochemical processes at the solid-liquid interface with electrochemical force microscopy. *Nature Communications* **5**, doi:Artn 3871

10.1038/Ncomms4871 (2014).
82  Yoo, H. I., Chang, M. W., Oh, T. S., Lee, C. E. & Becker, K. D. Electrocoloration and oxygen vacancy mobility of BaTiO3. *J Appl Phys* **102**, doi:Artn 093701

10.1063/1.2802290 (2007).
83  Cao, Y. & Kalinin, S. V. Phase-field modeling of chemical control of polarization stability and switching dynamics in ferroelectric thin films. *Phys Rev B* **94**, doi:Artn 235444

10.1103/Physrevb.94.235444 (2016).
84  Stephenson, G. B. & Highland, M. J. Equilibrium and stability of polarization in ultrathin ferroelectric films with ionic surface compensation. *Phys Rev B* **84**, 064107, doi:10.1103/PhysRevB.84.064107 (2011).
85  Strelcov, E. *et al.* Space- and Time-Resolved Mapping of Ionic Dynamic and Electroresistive Phenomena in Lateral Devices. *Acs Nano* **7**, 6806-6815, doi:10.1021/nn4017873 (2013).
86  Ding, J. L., Strelcov, E., Kalinin, S. V. & Bassiri-Gharb, N. Spatially Resolved Probing of Electrochemical Reactions via Energy Discovery Platforms. *Nano Lett* **15**, 3669-3676, doi:10.1021/acs.nanolett.5b01613 (2015).